\documentclass[showkeys,eqsecnum,showpacs,prd,superscriptaddress,nofootinbib]{revtex4}
\usepackage[dvips]{graphicx}
\usepackage{color}
\usepackage{amsmath}
\usepackage[dvips]{graphicx}
\begin{document}

\title{Phase Transition in Loop Quantum Gravity}

\author{Jarmo M\"akel\"a} 

\email[Electronic address: ]{jarmo.makela@vamk.fi}  
\affiliation{Vaasa University of Applied Sciences, Wolffintie 30, 65200 Vaasa, Finland}

\begin{abstract}  

We point out that with a specific counting of states loop quantum gravity implies that black holes perform a phase transition at a certain characteristic temperature $T_C$. In this phase transition the punctures of the spin network on the stretched horizon of the black hole jump, in effect, from the vacuum to the excited states. The characteristic temperature $T_C$ may be regarded as the lowest possible temperature of the hole. From the point of view of a distant observer at rest with respect to the hole the characteristic temperature $T_C$ corresponds to the Hawking temperature of the hole.

\end{abstract}


\maketitle

\section{Introduction}

Loop quantum gravity is a canonical quantum theory of gravitation, where the quantum states of the gravitational
field are the so-called spin network states. \cite{yy} In broad terms, spin network is a graph on a spacelike hypersurface of spacetime, where the time coordinate $t = constant$ such that every edge of the graph is associated with an irreducible representation of the group $SU(2)$. The most important prediction of loop quantum gravity is that the area eigenvalues of any spacelike two-surface of spacetime are of the form:
\begin{equation}
A = \gamma\ell_{Pl}^2\sum_p \sqrt{j_p(j_p+1)},
\end{equation}
where we have summed over the punctures $p$ of the spin network on the two-surface,
\begin{equation}
\ell_{Pl} := \sqrt{\frac{\hbar G}{c^3}}\approx 1.6 \times 10^{-35} m
\end{equation}
is the Planck length, and $\gamma$ is a pure number of the order of unity. The possible values taken by the quantum numbers $j_p$ are $0, \frac{1}{2}, 1, \frac{3}{2},...,$ and they determine the dimensions of the representation spaces of the group $SU(2)$ associated with the punctures $p$. Puncture $p$ is in vacuum, if $j_p = 0$; otherwise it is in an excited state.

   In a series of papers \cite{kaa,koo,nee} it has been shown that under the assumption that a specific stretched horizon of a black hole consists of discrete constituents, each of them contributing to the stretched horizon an area, which is an integer times a constant, the black hole performs, at a certain characteristic temperature $T_C$, a {\it phase transition}. In this phase transition the constituents of the stretched horizon jump, in effect, from the vacuum to the second excited states. The characteristic temperature $T_C$ corresponds, from the point of view of a distant observer at rest, to the Hawking temperature of the hole. \cite{vii} Since the constituents of the stretched horizon are in vacuum, when $T < T_C$, the Hawking temperature may be viewed, from the point of view of a distant observer at rest with respect to the hole, as the lowest possible temperature of the black hole.

   In this paper we shall show that a similar phase transition takes place in the black holes even, when the area spectra of their stretched horizons are of the form given in Eq. (1.1). According to the best knowledge of the author a phase transition of the same kind has not been reported before in loop quantum gravity. 

     Unless otherwise stated, we shall always use the natural units, where $\hbar = G = c = k_B = 1$.

\section{Partition Function}

   The thermodynamical properties of any system may be deduced from its partition function
\begin{equation}
Z(\beta) := \sum_n e^{-\beta E_n},
\end{equation}
where we have summed over the energy eigenstates $n$ of the system. $E_n$ is the energy eigenvalue associated with the energy eigenstate $n$. In Refs. \cite{kaa,koo,nee} the partition function of the black hole was written from the point of view of an observer on a stretched horizon, where the proper acceleration $a$ is a constant. In other words, no matter what may happen to the black hole, the proper acceleration $a$ is always kept unchanged. From the point of view of such observer the thermodynamics of black holes becomes very simple. For instance, it was shown in Refs. \cite{kaa,koo,nee} that the energy of the black hole takes, in the natural units, the form:
\begin{equation}
E = \frac{a}{8\pi}A,
\end{equation}
where $A$ is the area of the stretched horizon. Under the assumption that the spin network has $N$ punctures on the stretched horizon, and its area spectrum is given by Eq. (1.1), we find that the partition function of the black hole becomes to:
\begin{equation}
Z(\beta) = \sum_{n_1,n_2,...,n_N}\exp\left(-\beta T_0\sum_{p=1}^N\sqrt{n_p(n_p+2)}\right),
\end{equation}
where we have denoted:
\begin{equation}
T_0 := \frac{a}{16\pi}\gamma,
\end{equation}
and
\begin{equation}
n_p := 2j_p.
\end{equation}
Eq. (2.5) implies that the possible values of the quantum numbers $n_p$ are $0,1,2,...$.

      The question now is: In which way should we sum over the quantum numbers $n_p$? In what follows, we shall sum over the excited states of the punctures only. We shall also assume that when the quantum states of two different punctures are interchanged, the quantum state of the black hole is also changed, {\it i. e.} the punctures do not behave like bosons.  As a consequence, the partition function of the black hole takes, from the point of view of our observer, the form:
\begin{equation}
\begin{split}
Z(\beta) = &\sum_{n_1=1}^\infty\exp\left(-\beta T_0\sqrt{n_1(n_1+2)}\right)\\
                  &+\sum_{n_1=1}^\infty\sum_{n_2=1}^\infty \exp\left(-\beta T_0[\sqrt{n_1(n_1+2)} + \sqrt{n_2(n_2+2)}]\right)\\
                  &+\cdots\\
                  &+\sum_{n_1=1}^\infty\sum_{n_2=1}^\infty \cdots\sum_{n_N=1}^\infty\exp\left(-\beta T_0\sum_{p=1}^N\sqrt{n_p(n_p + 2)}\right).
\end{split}
\end{equation}
In the first term on the right-hand side of Eq. (2.6) just one puncture is in an excited state. In the second term two punctures are in the excited states, and so on. Finally, in the last term all $N$ punctures of the spin network on the stretched horizon are in the excited states. We may write Eq. (2.6) as:
\begin{equation}
Z(\beta) = \frac{1}{y} + \left(\frac{1}{y}\right)^2 + \cdots+ \left(\frac{1}{y}\right)^N,
\end{equation}
where we have denoted:
\begin{equation}
y = y(\beta) := \left[\sum_{n=1}^\infty\exp\left(-\beta T_0\sqrt{n(n+2)}\right)\right]^{-1}.
\end{equation}
On the right hand side of Eq. (2.7) we have a geometric series, and therefore:
\begin{equation}
Z(\beta) = \frac{1}{y - 1}\left[1 - \left(\frac{1}{y}\right)^N\right].
\end{equation}
Eq. (2.9) gives an expression for the partition function of the black hole from the point of view of an observer on the stretched horizon, where the proper acceleration $a = constant$. It is valid, whenever $y \ne 1$. In the special case, where $y = 1$, we have:
\begin{equation}
Z(\beta) = N.
\end{equation}

\section{Phase Transition}

  To investigate the phase transition taking place in the black hole, let us consider the average energy
\begin{equation}
E(\beta) = -\frac{\partial}{\partial\beta}\ln Z(\beta)
\end{equation}
of the hole. Using Eq. (2.9) we find:
\begin{equation}
E(\beta) = \left(\frac{1}{y-1} - \frac{N}{y^N - 1}\frac{1}{y}\right)\frac{dy}{d\beta}.
\end{equation}
We shall asume that the number $N$ of the punctures of the spin network on the stretched horizon is very large, indeed. Under this assumption we observe that something very interesting happens to the energy $E(\beta)$, when $y = 1$: When $y > 1$, the second term inside of the brackets in Eq. (3.2) vanishes in the limit, where $N\rightarrow \infty$, and hence we have, in effect,
\begin{equation}
E(\beta) = \frac{1}{y-1}\frac{dy}{d\beta},
\end{equation}
when $y > 1$. When $y < 1$, however, 
\begin{equation}
\lim_{N\rightarrow\infty}(y^N) = 0,
\end{equation}
and so we have, as an excellent approximation:
\begin{equation}
E(\beta) = \frac{N}{y}\frac{dy}{d\beta},
\end{equation}
whenever $y < 1$. So there is an enormous jump in the energy of the black hole when $y = 1$. Since $y$ depends on the temperature $T$ of the hole such that
\begin{equation}
y = \left[\sum_{n=1}^\infty\exp\left(-\frac{T_0}{T}\sqrt{n(n+2)}\right)\right]^{-1},
\end{equation}
this means that the black hole performs a {\it phase transition} at the characteristic temperature $T_C$, which is defined as a solution of the equation:
\begin{equation}
\sum_{n=1}^\infty\exp\left(-\frac{T_0}{T_C}\sqrt{n(n+2)}\right) = 1.
\end{equation}
When $T < T_C$, the punctures on the stretched horizon are, in effect, in vacuum, and there is no black hole. However, when $T > T_C$, the punctures suddenly jump to the excited states. In this sense, the characteristic temperature $T_C$ may be viewed as the lowest possible temperature of the hole. When the black hole radiates, the punctures of the spin network on its stretched horizon jump back to the vacuum, and radiation is emitted. In this sense loop quantum gravity, with its associated phase transition, provides a microscopic explanation to the Hawking effect.

   Unfortunately, it is impossible to solve the temperature $T_C$ exactly from Eq. (3.7). Nevertheless, if we denote by $x$ the solution of the equation
\begin{equation}
\sum_{n=1}^\infty \exp(-x\sqrt{n(n+2)}) = 1,
\end{equation}
then
 \begin{equation}
T_C = \frac{1}{x}T_0,
\end{equation}
and the numerical investigations imply:
\begin{equation}
x \approx 0.50764.
\end{equation}
Evaluating $\frac{dy}{d\beta}$ numerically, when $y = 1$, and employing Eq. (3.5)  one may obtain a numerical estimate for the jump $\Delta E$ in the energy during the phase transition:
\begin{equation}
\Delta E \approx 3.261NT_0.
\end{equation}
One may also show that the energy $E$ is an increasing function of the temperature $T$ of the black hole. Comparing Eqs. (2.3), (2.4), and (3.11) we observe that the average value taken by the quantum numbers $n_p$ after the phase transition is around 2.41.

    Using Eqs. (2.4) and (3.9) we find that if we choose the parameter $\gamma$ in Eq. (1.1) such that
\begin{equation}
\gamma = 8x \approx 4.061,
\end{equation}
then the characteristic temperature
\begin{equation}
T_C = \frac{a}{2\pi},
\end{equation}
which agrees with the {\it Unruh temperature} \cite{kuu} perceived by the observer on the stretched horizon, where the proper acceleration $a = constant$. In spherically symmetric spacetimes, where the line element takes, when written in the spherical coordinates $r$, $\theta$  and $\phi$, the form:
\begin{equation}
ds^2 = -f(r)\,dt^2 + \frac{dr^2}{f(r)} + r^2\,d\theta^2 +  r^2\sin^2(\theta)\,d\phi^2,
\end{equation}
the proper acceleration of an observer with constant coordinates $r$, $\theta$ and $\phi$ is: \cite{nee}
\begin{equation}
a = \frac{1}{2}f^{-1/2}\frac{df}{dr}.
\end{equation}
In Eqs. (3.14) and (3.15) $f(r)$ is a function of the radial coordinate $r$ only. In the Schwarzschild spacetime, for instance,
\begin{equation}
f(r) = 1 - \frac{2M}{r},
\end{equation}
where $M$ is the Schwarzschild mass of the hole we have:
\begin{equation}
a = \left(1 - \frac{2M}{r}\right)^{-1/2}\frac{M}{r^2}.
\end{equation}
Just outside of the event horizon of the black hole, where $r = 2M$, Eqs. (3.13) and (3.17) imply that the lowest possible temperature of the black hole from the point of view of our observer is:
\begin{equation}
T_C = \left(1 - \frac{2M}{r}\right)^{-1/2}\frac{1}{8\pi M}.
\end{equation}
According to the Tolman relation \cite{seite} the temperature $T_C$ corresponds, from the point of view of a faraway observer at rest, to the temperature
\begin{equation}
T_\infty = f^{1/2}T_C = \frac{1}{8\pi M},
\end{equation}
which agrees with the Hawking temperature \cite{vii}
\begin{equation}
T_H := \frac{1}{8\pi M}
\end{equation}
of the Schwarzschild black hole. So we have managed to show that according to loop quantum  gravity the Schwarzschild black hole has a certain minimum temperature which, with an appropriate choice of the parameter $\gamma$, agrees with the Hawking temperature of the hole from the point of view of a faraway observer. Even though we restricted our attention, for the sake of simplicity, to the Schwarzschild black hole, the same  result may be shown to hold for the other black holes as well. 

   Finally, we note in passing that the Bekenstein-Hawking entropy law \cite{vii,kasi} may also be obtained from loop quantum gravity in a very simple manner. The entropy of any system may be written, in the natural units, as:
\begin{equation}
S(\beta) = -\beta\frac{\partial}{\partial\beta}\ln z(\beta) + \ln Z(\beta),
\end{equation}
or if we use Eq. (3.1):
\begin{equation}
S(\beta) = \beta E(\beta)  + \ln Z(\beta).
\end{equation}
When $y < 1$, which means that $T > T_C$, Eq. (2.9) implies that we may write, in leading approximation for large $N$:
\begin{equation}
\ln Z(\beta) = N\ln\left(\frac{1}{y}\right),
\end{equation}
and therefore Eq. (3.22) takes the form:
\begin{equation}
S = \frac{a}{8\pi}\beta A + N\ln\left(\frac{1}{y}\right).
\end{equation}
Whe obtaining Eq. (3.24) we have used Eq. (2.2). In the limit, where $T \rightarrow T_C^+$, which means that $y\rightarrow 1^- $, we have:
\begin{equation}
S = \frac{1}{4}A, 
\end{equation}
which follows from the fact that $T_C = \frac{a}{2\pi}$. In the SI units we may write Eq. (3.25) as:
\begin{equation}
S = \frac{1}{4}\frac{k_Bc^3}{\hbar G}A.
\end{equation}
For all practical purposes we may identify the stretched horizon area $A$ with the event horizon area of the black hole. Equation (3.26) is exactly the same as the Bekenstein-Hawking entropy law, \cite{vii,kasi} which thereby has been obtained from loop quantum gravity. \cite{ysi}

\section{Effects of the Counting of States}

In addition to the counting of states employed in Eq. (2.6) in the calculation of the partition function, one may feel justified to use other kinds of countings of states as well. For instance, the dimension of the representation space of the group $SU(2)$  is $2j_p +1$, or $n_p +1$. This prompts us to introduce a $(n_p +1)$-fold degeneracy for every energy eigenstate associated with the given integer $n_p$. As a consequence, the partition function becomes to:
\begin{equation}
\begin{split}
Z(\beta) = &\sum_{n_1=1}^\infty (n_1 + 1)\exp\left(-\beta T_0\sqrt{n_1(n_1+2)}\right)\\
                  &+\sum_{n_1=1}^\infty\sum_{n_2=1}^\infty (n_1 + 1)(n_2 + 1)\exp\left(-\beta T_0[\sqrt{n_1(n_1+2)} + \sqrt{n_2(n_2+2)}]\right)\\
                  &+\cdots\\
                  &+\sum_{n_1=1}^\infty\sum_{n_2=1}^\infty \cdots\sum_{n_N=1}^\infty (n_1 + 1)(n_2 + 1)\cdots(n_N + 1)\exp\left(-\beta T_0\sum_{p=1}^N\sqrt{n_p(n_p + 2)}\right).
\end{split}
\end{equation}
Again, in the first term just one puncture is in an excited state, in the second term two punctures are in the excited states, and so on. As in Eqs. (2.7)  and (2.9) we may write the partition  function as:
\begin{equation}
Z(\beta) = \frac{1}{z} + \left(\frac{1}{z}\right)^2 + \cdots + \left(\frac{1}{z}\right)^N = \frac{1}{z - 1}\left[1  - \left(\frac{1}{z}\right)^N\right],
\end{equation}
but now we have:
\begin{equation}
z = z(\beta) := \left[\sum_{n=1}^\infty (n+1)\exp\left(-\beta T_0\sqrt{n(n+2)}\right)\right]^{-1}.
\end{equation}
Eq. (4.2) holds, whenever $z \ne 1$. If $z = 1$, we have $Z(\beta) = N$.

   A calculation similar to the one performed in Eqs. (3.1)-(3.5) indicates that there is still a phase transition, when $z = 1$, and the magnitude of the jump in the energy during this phase transition is:
\begin{equation}
\Delta E = N\frac{dz}{d\beta}.
\end{equation}
The characteristic temperature $T_C$ associated with the phase transition now satisfies an equation:
\begin{equation}
\sum_{n=1}^\infty(n+1)\exp\left(-\frac{T_0}{T_C}\sqrt{n(n+2)}\right) = 1,
\end{equation}
and the numerical investigations imply that
\begin{equation}
\frac{T_0}{T_C} \approx 0.861.
\end{equation}
The numerical investigations also imply that the magnitude of the jump in the energy during the phase transition is:
\begin{equation}
\Delta E \approx 2.921NT_0,
\end{equation}
and if we choose
\begin{equation}
\gamma = 8\frac{T_0}{T_C} \approx 6.888,
\end{equation}
the characteristic temperature $T_C$ agrees with the Unruh temperature $T_U$ perceived by our observer. Hence we find that the results produced by our modified counting of states are qualitatively similar to those produced by our original calculation. There are slight numerical differences, but otherwise the results are the same.

   It appears that the key point in the production of the phase transition in our calculation was an assumption that permutations of the nonvacuum quantum states of the punctures will always change the quantum state of the hole. As an example of a counting of states, which does {\it not} produce the phase transition, let us assume that the permutations of the nonvacuum states of the punctures do not change the quantum state of the hole. In this counting the partition function takes, when the $(2j_p + 1)$-fold degeneracy is taken into account, the form:
\begin{equation}
Z(\beta) = \frac{1}{z} + \frac{1}{2!}\left(\frac{1}{z}\right)^2 + \cdots + \frac{1}{N!}\left(\frac{1}{z}\right)^N,
\end{equation}
which follows from the fact that the number of the permutations of the quantum states of $N$ punctures is $N!$. In Eq. (4.9) $z = z(\beta)$ is the same as in Eq. (4.3). Since the total number $N$ of the puntures is assumed to be very large, we may write, in effect:
\begin{equation}
Z(\beta) = \exp\left(\frac{1}{z}\right) - 1.
\end{equation}
With this counting the average energy of the black hole is
\begin{equation}
E(\beta) = \frac{1}{z^2}\frac{\exp\left(\frac{1}{z}\right)}{\exp\left(\frac{1}{z}\right) - 1}\frac{dz}{d\beta},
\end{equation}
and there is no phase transition. We may thus conclude that the presence of the phase transition depends critically on whether the overall state of the black hole will change in the permutations of the quantum states of the punctures. If it does change, we have a phase transition, but if it does not, the phase transition does not exist, either.

\section{Concluding Remarks}

 In this paper we have pointed out that loop quantum gravity predicts an existence of a {\it phase transition} in black holes. During this phase transition the punctures of the spin network on the stretched horizon, where the proper acceleration $a = constant$ jump, in effect, from the vacuum to the excited states. From the point of view of a distant observer the phase transition temperature $T_C$ corresponds to the Hawking temperature of the hole. When $T < T_C$, the punctures are effectively in the vacuum, and there is no black hole. Hence the Hawking temperature may be viewed as the lowest possible temperature of the black hole.

    The results of this paper hinged on a specific counting of the states of the gravitational field, when obtaining an expression for the partition function of the black hole. More precisely, we summed over the nonvacuum states of the punctures only, and we assumed that if the quantum states of two punctures are interchanged, the state of the black hole is also changed. 

    Even though we have restricted our attention, in this paper, to the phase transition taking place in black holes, a similar phase transition may be shown to take place in the de Sitter spacetime as well. In the de Sitter spacetime the line element may be written as: \cite{kymppi}
\begin{equation}
ds^2 = -\left(1 - \frac{\Lambda}{3}r^2\right)\,dt^2 + \frac{dr^2}{1 - \frac{\Lambda}{3}r^2} + r^2\,d\theta^2 + r^2\sin^2(\theta)\,d\phi^2,
\end{equation}
The stretched horizon, where the proper aceleration $a = constant$ is replaced by a shrunken horizon just inside of the cosmological horizon, where
\begin{equation}
r = r_C := \sqrt{\frac{3}{\Lambda}}.
\end{equation}
As it was shown in Refs. \cite{nee,yytoo,kaatoo}, the energy of the de Sitter spacetime from the point of view of an observer on the shrinked horizon, where $a = constant$ is given by Eq. (2.2) with the stretched horizon area $A$ being replaced, in effect, by the area $A_C = 4\pi r_C^2 = {12\pi}/{\Lambda}$ of the cosmological horizon. Equation (3.11) implies that during the phase transition the cosmological constant $\Lambda$ decreases dramatically. The details of this phase transition are similar to those in Refs. \cite{nee,yytoo,kaatoo}. Choosing the number $N$ of the punctures of the spin network on the shrinked horizon to be around $10^{122}$ one finds that during the phase transition the cosmological constant drops from a huge Planck-size value to its presently observed value, which is around $10^{-35}s^{-2}$. Hence the phase transition in loop quantum gravity provides an explanation both to the presence and the smallness of the cosmological constant.

\begin{acknowledgments}

I am grateful to Aurelien Barrau for useful discussions and penetrating questions, which provided a starting point for this paper.

\end{acknowledgments}

\end{document}